\documentclass{nature}

\title{Biothermoeconomics analysis of cyanobacteria and microalga use for sustainable biofuel}

\usepackage[pdftex]{graphicx}
\usepackage{graphicx, floatflt}
\usepackage{graphicx, subfigure}

\author{Umberto Lucia$^{1,2}$ amd Giulia Grisolia$^{1,3}$}

\usepackage{multirow}
\usepackage{caption}
\usepackage{microtype}
\usepackage{amsmath}
\usepackage{amsfonts}
\usepackage{amssymb}
\usepackage{makeidx}
\usepackage{textcomp}
\usepackage{geometry}
\usepackage{microtype}
\usepackage{booktabs}
\usepackage{caption}
\usepackage{graphicx, floatflt}
\usepackage{subfig}
\usepackage{amsmath,amsthm,amssymb}
\usepackage{xcolor}
\begin{document}
	
	\maketitle
	
	\begin{affiliations}
		\item Dipartimento Energia \textquotedblleft Galileo Ferraris\textquotedblright, Politecnico di Torino, Corso Duca degli Abruzzi 24, 10129 Torino, Italy 
		\item umberto.lucia@polito.it
		\item giulia.grisolia@polito.it
	\end{affiliations}

	\begin{abstract}
Abstract\\Exergy is a thermodynamic quantity useful to obtain information on the work from any process. The analyses of irreversibility are important in the designing and development of the productive processes for the economic growth, but they play a fundamental role also in the analysis of socio-economic context. The link between the wasted exergy and the energy cost for maintain the productive processes are obtained in the bioengineering thermodynamics. This link holds to the fundamental role of fluxes and to the exergy exchanged in the interaction between the system and its environment. The equivalent wasted primary resource value is suggested as an indicator to support the economic considerations on the biofuel production by using biomass and bacteria. Moreover, the technological considerations can be developed by using the exergy inefficiency. Consequently, bacteria use can be compared with other way to obtain biofuels, by comparing both the technologies and the economic considerations.
	\end{abstract}
	
\section*{Introduction}
To date, our society continues to maintain a strong dependence upon fossil fuels as primary sources of energy, but now it has emerged the need of decreasing the emissions of the greenhouse gases.  For land-based transport electric power could represent an alternative technology to heat engine, but for aviation transport and shipping, up now, there exists no practical alternative engine in the foreseeable future. So, it is emerging a real interest in industrial development of liquid biofuels. Indeed, in the last decades, with reference to 1979s, a fundamental improvement of 30-40\% per annum in the production of biodiesel and bioethanol from crop plants has been carried out \cite{williams-laurens}, but they still continue to represent only less than 1\% of the world energy production: of the order of $10^{6}$ TJ for bioethanol in comparison of the order of around $10^{8}$ TJ of the global energy use.

Moreover, this first generation of biofuels has been the subject of a great number of ethical criticisms, both in relation to the land and water use, and to their conditioning of the food commodity price, with the related social consequences \cite{gallagher}. Consequently, a new improvement in the researches on new processes of biofuel production has occurred. Two different industrial processes are of interest:
\begin{itemize}
	\item the algal biomass production;
	\item the use of bacteria.
\end{itemize}

Macroalgae (seaweeds) have gained a place in the market and they are playing a growing role in biofuels production. From an energy production point of view, the fundamental difference between bacteria and eukaryotes is their subcellular structure; indeed, the former lack of the organelles (chloroplasts, mitochondria, nuclei), internal structures surrounded by lipid membranes whose bilayer structure requires strongly polar molecules. So, the major components of the cell membranes are just the high polar phospholipids and the glycolipids, which are the fundamental components of the algal lipids, with important industrial consequences \cite{williams-laurens}. Microalgae are able to photosynthesise, and they can live in many different habitats, from fresh to marine and hyper-saline environments \cite{falkowskiraven}. 

Today, a new growing sensibility in climate changes can represent a change in the viewpoint: microalgae are considered interesting in relation to the need of mitigating CO$_2$ release. Indeed, algal biomass can be used as a feedstock for the production of biodiesel, hydrogen, methane and bioethanol. Its production can represent a support to remove the carbon dioxide from the flue gases of fossil fuel. The growth rates of microorganisms can be very high because the metabolism of any living system is influenced by the surface-to-volume ratio: algae are able to divide once every 1–2 days up to every 3–4 h, under very  favourable conditions. Moreover, other advantages of microalgae as a feedstock for biofuel production, in relation to other biomass sources, must be considered:
\begin{itemize}
	\item there is no requirement for soil fertility;
	\item if marine algae are used, there is no need to draw upon supplies of freshwater.
\end{itemize}
Increasing the use of fuels from renewable biomass sources represents a fundamental opportunity both from an ecological point of view, towards a more sustainable energy system, and both an economic one. Recent political and research and development trends show a clear move towards lignocellulosic feedstocks for these biofuels; indeed, lignocellulosic  feedstocks represent a solution to mitigate the competition for land and water used for food production, by increasing the biomass production per unit of land area and reducing the inputs needed to grow the biomass itself \cite{tilman}.

Today, the fundamental key for improving the biofuel production systems consists in developing the efficient conversion technologies which are able to be economically competitive in relation to fossil fuels. The use of bacteria for lignocellulosic feedstocks could represents a possible solution. In this paper we wish to develop a thermodynamic and thermoeconomic analysis of this process.

Recently, an engineering thermodynamic approach has been developed \cite{lucia2012bioeng} by using a link between entropy generation principle and Constructal law. In this paper this thermodynamic is used to compare different biofuel production by using microalgae and cyanobacteria.

\section*{Results and Discussion}
Nowadays, growth is considered an imperative. In the analysis of the relation between energy and economic development it has been pointed out how development affects energy use, but it doesn't happen that the energy use affects development \cite{matoman}. But, it is clear that energy plays a fundamental role in promoting the economic growth \cite{stern1}, because energy is an essential factor of production because all economic processes require energy, while the economic analyses of growth are usually focused on the capital and labour.

Moreover, one of the main problems of industrialized countries is the management of CO$_2$ emissions. Economic strategy for the sustainable development suggest both to improve energy efficiency and to introduce a rational use of energy in all the member states of the European Union.

In order to evaluate the technological level, and the advanced level of industrial processes we introduce new indicators related to the inefficiency of the process and to the equivalent primary wasted resource value. This indicators allow us to link the exergy cost to the inefficiency of the system, allowing us to consider the cost of the wasted exergy for maintaining a process. 

Many factors can affect overall carbon dioxide emissions: economic growth levels, technological development, and production process selected for any particular production. At the same time, the CO$_2$ emission problem could represent a real opportunity to promote high-efficiency design of conventional plants, and consequent dissemination of advanced technologies. 

In Table \ref{lipidi} the lipid class distribution is summarised. Triglycerides, a storage lipid, may increase if the metabolic rate slows down. Consequently, lipid composition change during the different phases of growth. The cyanobacteria (prokaryotic algae) contains less total lipid content than the eukaryotic algae. the former, probably because the prokaryotes haven't got the internal membranes. So, the cyanobacteria potential as lipid producers doesn't appear to be promising, even if they are interesting for their easy structure due to their simple DNA, which characterise them as easier organisms for any industrial genetic manipulation.

In fuel production, the high concentration of unsaturated fatty acid is very important because it is a characteristics of the resultant fuel quality. In algae the unsaturated fatty acid needs seldom to be hydrogenated in order to improve the fuel properties because the unsaturated fatty acid increase the polymerisation in the engine oil with related problems with oxidative stability of the fuel.

The lignocellulosic biorefinery is based on a biochemical conversion platform where the polysaccharides in lignocellulosic material are converted to fuel by using enzymes \cite{huber}. One of the principal difficulty in the lignocellulosic biorefinery is the energy needs required by the process of fuel production \cite{ragauskas,cardona,piccolo}. Current integrated biorefinery involves the lignin-enriched residue as fuel for the energy needs, with the consequence of reducing the possibility of higher-value uses of lignin. The use of external sources of low-enthalpy heat can represent a way to overcome this unwanted consequence: waste heat from fermentation due to the presence of metabolism could be one of these sources.

Thermodynamic and thermoeconomic analysis, in particular the second law analysis, represents a powerful approach to evaluate the technological option for technology selection in order to design the more efficient biofuel production systems. In particular, our results can be summarized as follows:
\begin{itemize}
	\item for the \textit{Spirulina platensis}: $7.79$ J kg$^{-1}_{b}$
	\item for the \textit{Chlorella vulgaris}: $85.20$ J kg$^{-1}_{b}$
\end{itemize} 
where the indicator $EI_\lambda$ is expressed in kWh in order to assign an economic value comparable with other energy resources. The result is expressed in energy consumed per unit mass of biomass produced. In particular, considering the mean value of the kWh cost in Italy as 0.06797 EUR kWh$^{-1}$, obtaining the following cost of production:
\begin{itemize}
	\item for the \textit{Spirulina platensis}: 0.53 EUR kg$^{-1}_{b}$;
	\item for the \textit{Chlorella vulgaris}: 5.79 EUR kg$^{-1}_{b}$.
\end{itemize}
This result  highlight that biofuel becomes interesting from an economic point of view if obtained from cyanobacteria.

\section*{Methods}
Irreversible processes represent one of the fundamental topic of investigation in thermodynamic engineering for their fundamental role in the designing and development of the industrial devices and processes \cite{lucia-bioec-physica,lucia-bioec-gri}. 
Exergy is a quantity that allows the engineers to design system with the aim of obtain the highest efficiency at a least cost under the actual technology, economic and legal conditions, but also considering ethical, ecological and social consequences; indeed \cite{lucia-bioec-physica}:
\begin{enumerate}
	\item It allows the evaluation of the impact of energy resource utilization on the environment;
	\item	It allows the evaluation of more efficient energy-resource use, and of the locations, types, and magnitudes of wastes and losses;
	\item	It is an efficient technique to evaluate if it is possible to design more efficient energy systems by reducing the inefficiencies in existing technologies.
\end{enumerate}

Now, we wish to highlight that any effect in Nature is always the result of the dynamic balances of the interactions between the systems and their environment, so, the exchange of energy drives some behaviour of natural systems, i.e. their evolution is driven by the decrease of their free energy in the least time \cite{lucia-irrentrv,lucia-exfloconstr,bejanshape,bejaneghmf,bejanegm,bljhmt,blcl,bejanaet,annila1,annila2,annila4}.

In accordance with the first law of thermodynamics for open systems, any change in the energy of the system can be expressed in terms of the transfer of:
\begin{enumerate}
	\item Flows of matter across the system boundary, which bring internal, kinetic, chemical and other form of energy;
	\item Heat across the system boundary;
	\item Performance of work developed by or on the system.
\end{enumerate}
so, any process, interaction, cycle, etc. occur in a definite time $\tau$, which can be considered the lifetime of this phenomenon, and, in any process or interaction during this time, the energy variation $\Delta E$ of any open system results:
\begin{equation}
	\Delta E = \sum_i Q_i - W + \sum_j\int_0^\tau \dot{m}_i \big(h_j + e_{k,j}+e_{p,j}+e_{ch,j}\big)\,dt
\end{equation}
where $Q$ is the heat exchanges, $W$ is the work done, $\dot{m}$ is the mass flow, $h$ is the specific enthalpy, and $e$ is the specific energy, and the subscripts $k$, $p$ and $ch$ mean kinetic, potential and chemical respectively, $i$ and $j$ are related to the number of fluxes of heat and mass respectively. As a consequence of this energy variation, the following entropy variation, $\Delta S$, of the system occurs:
\begin{equation}
	\Delta S = \sum_i\frac{Q_i}{T_i} + \sum_j\int_0^\tau\dot{m}_i s_i dt + S_g
\end{equation}
where $T$ is the temperature of any $i-$th reservoir, $s$ is the specific entropy and $S_g = W_{\lambda} /T_0$ is the entropy variation due to irreversibility, named entropy generation \cite{lucia-irrentrv,lucia-exfloconstr,bejanshape,bejaneghmf,bejanegm,bljhmt,blcl,bejanaet} and $W_\lambda$ is the work lost.

Combining these equations, the following exergy balance can be obtained \cite{bejanaet}:
\begin{equation}
	W_t = \Delta B + \sum_{\alpha} J_{ex,\alpha} + \sum_{\beta} Ex_{Q,\beta} - T_0 S_g
\end{equation}
where:
\begin{itemize}
	\item $W_t$ is the net work done during the process;
	\item $\Delta B = E + p_0 V - T_0 S$ is the accumulation of non-flow exergy;
	\item $J_{ex} = \int_0^\tau\dot{m} \big(e - T_0 s\big) dt$ is the flow exergy due to mass flow;
	\item $Ex_{Q} = Q \big(1-T_0/T\big)$ is the exergy transfer due to heat transfer;
\end{itemize}
and the subscript $0$ means environment, while $p$ is the pressure and $V$ is the volume. The work lost $W_\lambda$ can be obtained as \cite{bljhmt}:
\begin{equation}
	W_\lambda = \frac{Ex_{in}-Ex_{out}-W}{T_0}
\end{equation}
where $Ex$ means exergy and $in$ and $out$ mean inflow and outflow respectively. So, the final relation useful for our analysis becomes:
\begin{equation}
	\begin{aligned}
		T_0 S_g &= \sum_j\int_0^\tau \dot{m}_i \big(h_j + e_{k,j}+e_{p,j}+e_{ch,j}\big)\,dt + \sum_\ell\int_0^\tau \dot{n}_{\ell}\nu_{\ell} \big(g_{\ell}^\oplus - ex_{ch,\ell}^\oplus\big)\,dt - \\ &- \sum_i \Bigg(1-\frac{T_0}{T}\Bigg)\,Q - W_t - \int_0^\tau \frac{d}{dt}\big(E-T_0 S\big)dt
	\end{aligned}
\end{equation}
where $g$ is the molar specifis Gibbs potential, $ex_{ch} = y \big(\mu - \mu_0\big)_{T_0,p_0}$ is the molar specific chemical exergy at the reference atmosphere, $y$ is the molar fraction, $\dot{n}$ is the molar flus, $\nu$ is the stoichiometric coefficient,  and $\mu$ is the chemical potential, $\oplus$ means means standard conditions.

Consequently, it is a fundamental quantity, useful both for engineering application of living structures, as bacteria use in fermentation, etc., and in biological or medical study of evolution of diseases.

Photosynthesis is a process complex organic molecules result from simple molecules by absorbing solar radiation \cite{zavala,andriesse}. The most studied chemical reaction is the one made by superior plants and bacteria \cite{zavala}:
\begin{equation}
	6 \text{CO}_2 + 6 \text{H}_2\text{O} \rightarrow \text{C}_6\text{H}_{12}\text{O}_6 + 6 \text{O}_2
\end{equation}
Other reactions are done by:
\begin{enumerate}
	\item superior plants and cyanobacteria:
	\begin{equation}
		\label{11}
		6 \text{CO}_2 + 12 \text{H}_2\text{O} \rightarrow \text{C}_6\text{H}_{12}\text{O}_6 + 6 \text{H}_2\text{O} + 6 \text{O}_2
	\end{equation}
	\item sulfur purple bacteria and sulfur green bacteria, young bacteria:
	\begin{equation}
		\label{12}
		6 \text{CO}_2 + 12 \text{H}_2\text{S} \rightarrow \text{C}_6\text{H}_{12}\text{O}_6 + 6 \text{H}_2\text{O} + 12 \text{S}
	\end{equation}
	\item sulfur purple bacteria, old bacteria:
	\begin{equation}
		\label{13}
		6 \text{CO}_2 + 6 \text{H}_2\text{O} + 3 \text{H}_2\text{S} \rightarrow \text{C}_6\text{H}_{12}\text{O}_6 + 3 \text{H}_2\text{SO}_4
	\end{equation}
	\item sulfur purple bacteria and sulfur green bacteria:
	\begin{equation}
		\label{14}
		6 \text{CO}_2 + 15 \text{H}_2\text{O} + 3 \text{Na}_2\text{S}_2\text{O}_3 \rightarrow \text{C}_6\text{H}_{12}\text{O}_6 + 6 \text{H}_2\text{O} + 6 \text{NaHSO}_4
	\end{equation}
	\item non sulfur purple bacteria and non sulfur green bacteria:
	\begin{equation}
		\label{15}
		6 \text{CO}_2 + 12 \text{CH}_3\text{CH}_2\text{OH} \rightarrow \text{C}_6\text{H}_{12}\text{O}_6 + 12 \text{CH}_3\text{CH=} + 6 \text{H}_2\text{O}
	\end{equation}
	\item non sulfur purple bacteria and non sulfur green bacteria:
	\begin{equation}
		\label{16}
		2 \text{CO}_2 + 4 \text{CH}_3\text{OH} \rightarrow \text{C}_6\text{H}_{12}\text{O}_6 + 2 \text{H}_2\text{O}
	\end{equation}
	\item non sulfur purple bacteria:
	\begin{equation}
		\label{17}
		6 \text{CO}_2 + 12 \text{Succinic acid} \rightarrow \text{C}_6\text{H}_{12}\text{O}_6 + 12 \text{Fumaric acid}
	\end{equation}
	\item non sulfur purple bacteria:
	\begin{equation}
		\label{18}
		6 \text{CO}_2 + 12 \text{Malic acid} \rightarrow \text{C}_6\text{H}_{12}\text{O}_6 + 12 \text{Oxalacetic acid}
	\end{equation}
	\item heliobacteria:
	\begin{equation}
		\label{19}
		\begin{aligned}
			3 \text{CH}_3\text{COOH} + 6 \text{H}_2\text{O} \rightarrow 6 \text{CO}_2 + 12 \text{H}_2
			\\
			6 \text{CO}_2 + 12 \text{H}_2 \rightarrow \text{C}_6\text{H}_{12}\text{O}_6 + 6 \text{H}_2\text{O}
		\end{aligned}
	\end{equation}
\end{enumerate}

In order to use the entropy generation principle for an engineering thermodynamic analysis of photosynthesis it is necessary to evaluate the entropy generation. To do so the system must be defined and the fluxes must be evaluated. The Sun, the photosynthetic organism and the Earth are three different systems \cite{zavala}, from which there exist flows. The process considered can be analysed following three steps:
\begin{enumerate}
	\item light comes from sun to the photosynthetic organism without any work carrying an energy (and exergy) flux. A gas of photons is emitted from the Sun. In the path from the Sun to the Earth this gas of photon follows an adiabatic expansion, with a consequent dilution of photons. Consequently, the Sun can be considered as a grey-body at temperature $T_S = 5762$ K in radiative equilibrium with the Earth, which absorbs all the radiation, so the Earth can be considered as a black body at atmospheric temperature $T_E = 298.15$ K. The energy balance results:
	\begin{equation}
		\begin{aligned}
			\varepsilon \sigma T_S^4 = \sigma T_E^4
			\\
			\varepsilon = \frac{R_S^2}{R_O^2}
		\end{aligned}
	\end{equation}
	with $\sigma = 5.67\times 10^{-8}$ Wm$^{-2}$K$^4$, $\epsilon$ emissivity, $R_S$ Sun radius and $R_O$ Earth radius. The entropy generation during the process can be evaluated as \cite{zavala}:
	\begin{equation}
		S_{g,SE} = \frac{4}{3}\,60 N_A h\nu \,\bigg(\frac{1}{T_E}-\frac{1}{T_S}\bigg)
	\end{equation}
	with $\nu =c/\lambda$ frequency, $c$ light velocity and $\lambda$ wave length, $h = 6.626\times 10^{-32}$ Js Planck's constant and $N_A = 6.022\times 10^{23}$ mol$^{-1}$;
	\item the photosynthetic organism absorbs the light from the environment, of which the entropy generation is $S_{g,la} = 0$ J K$^{-1}$, because it happens at constant temperature ($T_{PO} = T_E$, where PO means photosynthetic organism) without any work;
	\item the photosynthetic organism produces glucose using the exergy absorbed from the light with an entropy generation:
	\begin{equation}
		S_{g,gp} = -\frac{\Delta G_{PO}}{T_{PO}}
	\end{equation}
	where PO means photosynthetic organism;
	\item the photosynthetic organism exchange the remaining heat with Earth, of which the entropy generation $S_{g,POE} = 0$ J K$^{-1}$ because it happens at the same temperature without any work.
\end{enumerate} 
As a consequence the total entropy generation for the photosynthesis process results:
\begin{equation}
	S_{g,PS} = S_{g,SE} + S_{g,la} + S_{g,gp} + S_{g,POE} = \frac{4}{3}\,60 N_A h\nu \,\bigg(\frac{1}{T_E}-\frac{1}{T_S}\bigg) - \frac{\Delta G_{PO}}{T_{PO}}
\end{equation}
The consequent reaction efficiency can be evaluated as:
\begin{equation}
	\eta = \frac{\Delta G^0}{60 N_A h\nu}
\end{equation}
For the above nine chemical reaction, the entropy generation is evaluated in Table \ref{tabella}. It can be pointed out that from an engineering approach the superior plants have the higher efficiency.

This last relation allows us to evaluate all the dissipations during the process, and to define a parameter useful to quantify the technological level of a process related just to the unavailability, named exergy inefficiency, as \cite{lucia-bioec-physica,lucia-bioec-gri}:
\begin{equation}
	\varepsilon_\lambda = \frac{T_0 S_g}{Ex_{in}}
\end{equation}
which is useful to evaluate the technological maturity of a production system or a production sector in a country, because it allows us to obtain information on the losses of processes. The less is the value of the unavailability percentage the more the industrial process is efficient in terms of energy use \cite{Lucia-exergyRSER,lucia-bioec-physica,lucia-bioec-gri}.

Moreover, starting from these results, we can also define the sustainability of a process, by using a new indicator, the equivalent wasted primary resource value for the work-hour, defined as:
\begin{equation}
	\label{equazioneA}
	EI_\lambda = \frac{T_0 S_g}{n_h n_w}
\end{equation}
where $n_h$ is the working hours and $n_w$ is the number of workers. This quantity allows us to quantify the cost of the wasted exergy necessary to support the workhours and to generate capital flows.

In order to show the use of these two last relations we compare:
\begin{itemize}
	\item the cyanobacterium \textit{Arthrospira platensis}, known as \textit{Spirulina platensis};
	\item the microalga \textit{Chlorella vulgaris};
\end{itemize} 
with the aim to develop thermoeconomic considerations on the biomass obtained. The properties of these biosystems are summarized in Table \ref{clorespiru}. We have no data on a possible number of worker and hour of worker for an industrial use of these biosystems, so, in order to develop our evaluation, we modify the relation (\ref{equazioneA}) as follows:
\begin{equation}
	EI_\lambda = \frac{T_0 s_{g,PS}\, n_{\text{CO}_2}}{\dot{m}_B}
\end{equation}
where $\dot{m}_B$ is the biomass produced in a day and $n_{\text{CO}_2}$ is the mole of CO$_2$ used by the biosystem. The two biosystems has the same metabolic properties represented by the chemical equation (\ref{11}). The temperature $T_0$ considered is the temperature of water in a photobioreactor, 30\textcelsius $\,$ (= 300.15 K).

%	\begin{figure}
%	\caption{Expansion of the universe. The thermodynamic system is the volume of universe between two different position generated by the universe expansion. A volume element of mass flows between these two positions.}
%	\includegraphics{Figure1}
%\end{figure}

	%% Put the bibliography here, most people will use BiBTeX in
	%% which case the environment below should be replaced with
	%% the \bibliography{} command.
	
		\section*{References}
	\bibliography{References}

%	\begin{thebibliography}{1}
%		\bibitem{dummy} Articles are restricted to 50 references, Letters
%		to 30.
%		\bibitem{dummyb} No compound references -- only one source per
%		reference.
%	\end{thebibliography}

	%% Here is the endmatter stuff: Supplementary Info, etc.
	%% Use \item's to separate, default label is "Acknowledgements"
	\newpage
	\begin{addendum}
%		\item Put acknowledgements here.
		\item[Competing Interests] The authors declare that they have no
		competing financial interests.
			\item[Contributions] UL developed the thermal engineering model. GG developed the numerical evaluation. UL wrote the paper. All the authors read and correct the paper.
		\item[Correspondence] Correspondence and requests for materials
		should be addressed to Umberto Lucia, Dipartimento Energia \textquotedblleft Galileo Ferraris\textquotedblright, Politecnico di Torino, Corso Duca degli Abruzzi 24, 10129 Torino, Italy ~(email: umberto.lucia@polito.it).
	\end{addendum}
	
	%%
	%% TABLES	
	%%
	%% If there are any tables, put them here.
	%%
	\newpage
	\begin{table}
		\caption{Mean lipid content as a percent of total lipid \cite{williams-laurens,Borowitzka}}	
		\centering
		\begin{tabular}{l c c c}
			\hline
			\label{lipidi}
			& & & \\
			Algae species & Simple lipids & Glycolipids & Phospholipids \\
			& & & \\
			\hline 	
			& & & \\
			Chaetoceros species & 37 -- 16 & 36 -- 8 & 25 -- 8 \\
			Phaeodactylum tricornutum & 54 -- 6 & 34--5 & 11--1 \\
			Chlamydomonas species & 48 -- 10 & 44 -- 13 & 6 -- 3 \\
			Dunaliella tertiolecta & 7 -- 1 & 67--1 & 25--0 \\
			Dunalliella viridis & 13--1 & 44--3 & 42--2 \\
			Nannochloropsis oculata & 22 -- 1 & 39--0 & 38--1 \\
			Isochrysis species & 36--3 & 35--1 & 27--3\\
			& & & \\
			\hline
		\end{tabular}
	\end{table}
	
	\newpage
	\begin{table}
		\caption{Entropy generation and efficiency for photosynthesis \cite{zavala}}\label{tabella}
		\centering
		\begin{tabular}{|c|c|c|c|c|}
			\hline
			$\lambda$ & reaction & $\Delta G^0$  & $\Delta s_{g,PS}$ & $\eta$ \\
			\hline
			$[$nm$]$ & & [kJ mol$^{-1}$]  & [kJ mol$^{-1}$K$^{-1}$] & \% \\  \hline
			680 & (\ref{11}) & 2880.31 & 37.543 & 27.288 \\ \hline
			798 & (\ref{19}) &320.65 & 39.148 & 3.565 \\ \hline
			\multirow{4}*{840} & (\ref{12}) & 429.64 & 36.771 & 5.028 \\ %\cline{2-5}
			& (\ref{14}) & 621.47 & 36.128 & 7.273\\ %\cline{2-5}
			& (\ref{15}) & 584.86 & 36.251 & 6.845 \\ %\cline{2-5}
			& (\ref{16}) & 71.27 & 37.973 & 0.834 \\ \hline
			\multirow{7}*{870} & (\ref{12}) & 429.64 & 35.454 & 5.208 \\ %\cline{2-5}
			& (\ref{13}) & 744.57 & 34.397 & 9.025 \\
			& (\ref{14}) & 621.47 & 34.810 & 7.533\\ %\cline{2-5}
			& (\ref{15}) & 584.86 & 34.933 & 7.089 \\ %\cline{2-5}
			& (\ref{16}) & 71.27 & 33.197 & 0.864 \\ 
			& (\ref{17}) & 1066.56 & 33.137 & 12.928 \\
			& (\ref{18}) & 609.48 & 34.850 & 7.388 \\ \hline
			\multirow{3}*{890} & (\ref{12}) & 429.64 & 34.625 & 5.327 \\ %\cline{2-5}
			& (\ref{13}) & 744.57 & 33.568 & 9.232 \\
			& (\ref{14}) & 621.47 & 33.981 & 7.706\\ \hline
			\multirow{4}*{960} & (\ref{15}) & 584.86 & 31.474 & 7.822 \\ %\cline{2-5}
			& (\ref{16}) & 71.27 & 33.197 & 0.953 \\
			& (\ref{17}) & 1066.56 & 29.859 & 14.265\\ 
			& (\ref{18}) & 609.48 & 31.392 & 8.152 \\ \hline
		\end{tabular}
	\end{table}

	\newpage
	\begin{table}
		\caption{Data for numerical evaluation \cite{singh-singh}}	
		\centering
		\begin{tabular}{l c c}
			\hline
	\label{clorespiru}
			& & \\
			Biosystem & Biomass produced & CO$_2$ fixed \\
			& [kg m$^{-3}$d$^{-1}$] & [$10^{-3}$ kg m$^{-3}$d$^{-1}$] \\ 
			& & \\
			\hline 	
			& & \\
			\textit{Spirulina platensis} & 2.91 & 318.61 \\
			\textit{Chlorella vulgaris} & 0.21  &  251.64 \\
			& & \\
			\hline
		\end{tabular}
	\end{table}

\end{document}